\documentstyle[12pt,aaspp4]{article}

\begin{document}

\title{Orbits of Globular Clusters in the Outer Galaxy: NGC 7006}
\author{Dana I. Dinescu\altaffilmark{1,2}, Steven R. Majewski\altaffilmark{1,3}, Terrence M. Girard\altaffilmark{4} and 
Kyle M. Cudworth\altaffilmark{5}}

\altaffiltext{1}{Department of Astronomy, University of Virginia, Charlottesville, VA 22903-0818 (dd7a@virginia.edu, srm4n@virginia.edu)}
\altaffiltext{2}{Astronomical Institute of the Romanian Academy, Str.
Cutitul de Argint 5, RO-75212, Bucharest 28, Romania}
\altaffiltext{3}{David and Lucile Packard Foundation Fellow, Cottrell Scholar
of the Research Corporation} 
\altaffiltext{4}{Astronomy Department, Yale University, P.O. Box 208101,
New Haven, CT 06520-8101 (girard@astro.yale.edu)}
\altaffiltext{5}{Yerkes Observatory, 373 W. Geneva St.
Williams Bay, WI 53191-0258 (kmc@yerkes.uchicago.edu)}

\begin{abstract}
We present a proper motion study of the distant
globular cluster NGC 7006 based on the measurement of 25 photographic
plates spanning a 40-year interval.

The absolute proper motion determined with respect to extragalactic objects is
($\mu_{\alpha}$ cos $\delta$, $\mu_{\delta}$) = (-0.96, -1.14) $\pm$
(0.35, 0.40) mas yr$^{-1}$. The total space velocity of NGC 7006
in a Galactocentric rest frame is 279 km s$^{-1}$,
placing the cluster on one of the most energetic orbits
($R_{a}$ = 102 kpc)  known to date
for clusters within 40-kpc from the Galactic center.

We compare the orbits --- as determined from full space velocities ---
of four clusters that have apocentric 
radii larger than 80 kpc with those of Galactic satellites with
well-measured proper motions. These clusters are NGC 5466, 
NGC 6934, NGC 7006 and Pal 13 and the satellites are 
the Sagittarius dwarf spheroidal galaxy (dSph), the Large Magellanic
Cloud, Ursa Minor dSph and Sculptor dSph.
Only NGC 5466 and NGC 6934 seem to have similar orbital 
parameters, indicating a possible phase-space association.
NGC 7006, Pal 13 and the ``pair'' NGC 5466, NGC 6934 
do not show any dynamical association with the Galactic satellites
considered here.
NGC 5466, NGC 6934, NGC 7006 and  Pal 13 have orbits which 
are highly eccentric and of various inclinations 
with respect to the Galactic plane. In contrast, 
the orbits of the Galactic satellites 
are of low to moderate eccentricity and highly inclined.
Based on orbit types, chemical abundances
and cluster parameters, we discuss the properties of 
the hypothetical host systems 
of the remote globulars in the Searle-Zinn paradigm.
It is apparent that clusters such as NGC 5466, NGC 6934 and NGC 7006 formed in 
systems that more likely resemble the Fornax dSph, rather than the
Sagittarius dSph. 

We also discuss plausible causes for the difference found so far
between the orbit type of outer halo clusters and that of Galactic
satellites and for the tentative, yet suggestive
phase-space scatter found among outer halo clusters.

\end{abstract}


\section{Introduction}

Since the early study of Sandage \& Wildey (1967), NGC 7006
(C2059+160, $l = 63\fdg8$, $b = -19\fdg4$)
has been known to be a globular cluster with an unusual red horizontal branch 
(HB) for its low metallicity ([Fe/H] = -1.63, Harris 1996 \footnote[6]
{http://physun.physics.mcmaster.ca/Globular.html}, hereafter H96).
As an archetypical ``second-parameter'' cluster, and residing in 
the outskirts of the Milky Way at $\sim 37$ kpc from the Galactic center
(GC), NGC 7006 is of considerable interest for
formation scenarios of the assembling of the Galactic halo.
Specifically, its age can constrain timescales of the assembling 
process, while the shape of the orbit can bring some insights
as to how this process proceeded.
Progress in determining the age of NGC 7006 has already been made
by Buonanno {\it et al.} (1991), hereafter B91. From the analysis of the
main sequence turnoff, B91 find no evidence that NGC 7006 is significantly
younger than the bulk of the globular clusters. We will come back to this 
issue in our Discussion Section. 

NGC 7006 is located very close to the radius where the
the globular-cluster spatial distribution is
truncated ($R_{GC} \sim 40$ kpc; from H96 there are 
no clusters between 40 and 70 kpc and only 6 clusters
beyond 70 kpc; see also Zinn 1988). Yet, indications
are that it is not at apogalacticon, as evidenced by the cluster's
large radial velocity
( -384 km s$^{-1}$ H96).  A cluster  likely 
to have excursions into the remote regions of the Galaxy is
of interest because it may be dynamically associated with some of the Galactic
satellite galaxies. 
Possible dynamical associations of some remote clusters with satellites
based on radial velocities and locations
have been proposed and studied already
(e.g., Lynden-Bell \& Lynden-Bell 1995; Palma, Majewski \& Johnston 2001).
However, in order to quantify such an association reliably one 
needs tangential velocities; it is the purpose of the present work to
determine them for NGC 7006 with the plate collection that we have available
 for this cluster. This study continues
a program (Majewski \& Cudworth 1993) to derive proper motions
of a number of distant globular clusters and dwarf spheroidals (dSph).

The paper is organized as follows: in Section 2 we describe the
photographic photometry and astrometry, in Section 3
we determine the absolute proper motion based on extragalactic
objects, in Section 4 we derive the orbit of NGC 7006, and in
Section 5 we discuss the cluster's orbit in relation to orbits of
other clusters and satellites that have well-measured absolute proper motions.
Finally, in Section 6 we summarize our results.

\section{Observations and Measurements}

The photographic plate material available for this study, as well as
the reduction procedure are very similar to those used in our
previous paper on the cluster Pal 12 (Dinescu {\it et al.} 2000, 
hereafter D2K). Therefore we refer the reader to that paper for 
a detailed description of the procedures used. Table 1 summarizes 
the collection of 25 plates used in this study: 10 were taken with 
the Las Campanas DuPont 2.5 m reflector
(scale = $10\farcs92$ mm$^{-1}$), 7 with 
the KPNO 4 m reflector (scale = $18\farcs60$ mm$^{-1}$
UBK7 corrector), and 8 with the Hale 5 m reflector (scale =
 $11\farcs12$ mm$^{-1}$). 
An input catalog containing 5696 objects
was prepared from a full digitization of
plate CD3062, which is one of the deepest, good-quality plates among the
modern-epoch set (Table 1).  The digitized area is a square of $26\arcmin$
(143 mm for the DuPont plate scale) on a side.
The digitization was done with the
University of Virginia
PDS microdensitometer (30-$\mu$m pixel 
size), and preliminary positions, object diameters and
object classification  were determined using the
FOCAS software\footnote[7]{ftp://iraf/noao.edu/iraf/docs/focas/focasguide.ps}
 (Valdes 1982, 1993). 
Using this input catalog, all of the
plates were measured in a fine-raster, object-by-object
mode with the Yale PDS microdensitometer, using a 12.7-$\mu$m pixel size for 
the DuPont 2.5 m and Hale 5 m plates,
and a 10-$\mu$m pixel size for the KPNO 4 m plates.
The image positions, instrumental magnitudes and image parameters
were determined using the Yale Image Centering routines
(two-dimensional, bivariate Gaussian fit, Lee \& van Altena 1983). 
Due to the thermal drift in the PDS during long scans,  
eight stars were repeatedly measured in order to monitor and correct for 
drifts in the measurement system. This correction includes terms for 
translation and rotation.
The image-centering accuracy for well-measured, stellar objects
ranges between 0.8 and 1.7 $\mu$m, depending on the plate emulsion.

\subsection{Photographic Photometry}

The photographic photometry was calibrated using sequences 
obtained from CCD $BV$ photometry from B91,
their field 2, that includes their best-quality CCD photometry, and
photoelectric photometry from 
Sandage \& Wildey (1967) to ensure a good calibration at the bright end
($V \le 15.5$). We have compared the B91 CCD photometry with the 
photoelectric photometry from Sandage \& Wildey (1967) using four 
stars that were measured in both studies. We have found an offset of
0.1 mags in $V$ and 0.2 mags in $B$ between the two studies. 
We have then applied these offsets to the Sandage \& Wildey (1967)
photometry, such that the calibrating photometry is now on the system 
of B91.

For the $B$ band we have used only the modern-epoch plates (8 plates, see
Table 1) taken with the DuPont 2.5 m reflector, while for the $V$ band
we have used 2  modern-epoch plates and 3 intermediate-epoch plates
taken with the KPNO 4 m reflector.
Each plate requires a separate calibration and that calibration is determined
via cubic spline interpolation to a calibration sequence.
The final calibrated magnitude for each star is determined
from the average of the measurements from each plate and
the error in the magnitude is given by the scatter of
the measurements, after outliers have been eliminated.
From this scatter, we obtain an error of 0.03 in $B$ and
0.05 in $V$ for stars brighter than $V = 19$. For fainter magnitudes,
the errors increase rapidly. The $B-V$ color is obtained from
the straight differences of the averaged $B$ and $V$ magnitudes.
The formal error in the $B-V$ color is of the order of 0.06 mag for 
well-measured stars ($ V \le 19$).
A direct comparison of the calibrated magnitudes with the 
standard ones gives somewhat larger errors in the magnitudes:
0.05 in $B$, 0.09 in $V$ and $\sim 0.12$ in $B-V$. This is
because the calibrating sequences are determined primarily from
standards located in a  region
centered on the cluster, where crowding effects increase the
errors in the instrumental photographic magnitudes.

The photometry determined here was used in our 
astrometric reduction described below (Section 2).

\subsection{Astrometry}

All plates were precorrected for distortion as modeled by
Cudworth \& Rees (1991) for the DuPont 2.5 m 
and Chiu (1976) for the KPNO 4 m and Hale 5 m.
The DuPont 2.5 m plates were also precorrected for differential refraction
since, for this telescope, the differential refraction correction 
is comparable with the distortion correction (see D2K).
In what follows all the linear
dimensions  correspond to the DuPont plate scale (see Table 1).

Magnitude-dependent systematics were modeled from the 
residuals given by the cluster stars, where the preliminary list of cluster
stars is defined by positions in the color-magnitude diagram (CMD)
(see Guo {\it et al.} 1993, Dinescu {\it et al.} 1996).
The plate transformation included polynomials of up to fourth-order terms and
linear color terms. As in our previous paper
(D2K), we found significant color terms for both the
KPNO 4 m and Hale 5 m plates (up to $5~\mu$m mag$^{-1}$), with the largest
terms in the y-coordinate (aligned with declination).
Since the Hale 5 m plates have the largest 
systematics among our three sets of plates, and the 
image quality degrades rapidly with distance from the plate center
(D2K, Siegel {\it et al.} 2001),
we have firstly determined
preliminary proper motions from the DuPont 2.5 m and the KPNO 4 m
plates, using  an iterative 
central-plate-overlap algorithm (see for instance Girard {\it et al.} 1989).
Then, with these preliminary proper motions, we modeled the 
Hale 5 m plates, that were afterwards introduced into the solution.

The proper motion is calculated for each star from a linear least squares 
fit of positions as a function of plate epoch. The error in the proper motion
is given from the scatter about this best-fit line. Measurements that
differ by more than $0\farcs2$ (18 $\mu$m)
 from the  best-fit line were excluded.

In Figure 1, top panels, we show the proper-motion error in each
coordinate as a function
of magnitude for stars that had at least six measurements and that
reside within a $14\farcm6$ box centered on the 
plate center. Outside of this box, the Hale 5 m measurements degrade
quickly and objects tend to have measurements only from the two
more modern set of plates.
Well-measured stars 
($ 16 < V < 18 $) have a mean internal proper-motion error
of 0.15 mas yr$^{-1}$ in each coordinate. For these stars,
the mean internal positional error at the mean epoch of 1977.0
is 2.7 mas (0.25 $\mu$m) in each coordinate.
The middle panels show the the proper motions as a function of 
$V$ magnitude, while the bottom panels show the proper motions as a
function of $B-V$ color. We have highlighted the cluster stars
as defined by the CMD of B91.
A larger scatter is present in the $y$ coordinate, which is
mainly due to residual color terms in this direction. However,
within the uncertainties, there are no significant magnitude
and color trends in the proper motions over the range (Fig.1) defined
by cluster stars.

The present study is complete only to $V \sim 19$ due to the shorter 
exposure times for this plate collection compared to that for the 
study of Pal 12 (D2K). Proper motion membership probabilities
were also calculated, and they will be presented elsewhere.

\section{Absolute Proper Motion}

\subsection{The Relative Motion of the Galaxies} 

We have identified the galaxies to be used as references for
the absolute proper motion  based on the distribution of 
image parameters such as peak density versus radius and instrumental magnitude.
The image parameters for this selection are those derived from
the KPNO 4 m plate 2874 (Table 1), which has a fine-grained emulsion.
Each potential galaxy thus selected was visually inspected
to ensure that blended and other spurious objects are not included 
in the list. Twenty six  galaxies were so identified and, from these, we
discarded those that had magnitudes and colors far outside the range 
of the cluster stars (Fig. 1). We were left with 19 galaxies,
spread across the entire area measured (a square of 
25 $\arcmin$ on a side). In addition, one known QSO resides in the field
of NGC 7006: QSO2059+1604 (Harris {\it et al.} 1992).

As in our previous paper (D2K), position-dependent systematics 
are present in the proper motions as the distance from the plate
center increases. These systematics are left
because of our inability to determine a plate model accurately
in the outer regions of the field, where both distortion and coma 
increase considerably for the Hale 5 m plates.
Siegel {\it et al.} (2001) in their photographically
similar study of Pal 13, restricted the usable area for astrometry
to a circle of $5\farcm5$ radius (30 mm). Within this area
they had 16 extragalactic objects. Unfortunately, the NGC 7006 field has
only four extragalactic objects within a 30-mm radius from the plate center.
Since galaxies have a poorer centering accuracy than do stars, due to
their more shallow sloping profiles, the use of only
four galaxies would not provide a suitably accurate
calibration of the correction needed to yield an absolute proper motion.

In order to be able to use all 20 of the extragalactic
objects, we apply the local-solution 
method developed in D2K and Dinescu {\it et al.} (1997).
For each galaxy a local reference system is defined and
the proper motion of the galaxy is re-determined with respect to this
local system. The assumption is that, locally, both the galaxy
and its reference system are affected by the same amount of
geometric systematics and, thus, when referring one to the other,
the systematics cancel out.
The local reference system is chosen from field stars in 
a given magnitude range. 
The number of reference stars within a local system
is chosen such that the area occupied is small enough that 
geometric systematics are unimportant and large enough that 
the sample size minimizes the scatter due to the
intrinsic proper-motion dispersion of field stars in that
particular magnitude range.
For each local system, we have chosen 20 field stars
with $ 16 < V < 19$; the radii of these local systems range
between $0\farcm7$ and $1\farcm5$ (4-8 mm). 
The mean motion of this local reference system
is defined by the median value of the proper motions
and this value is  subtracted from the motion of the galaxy. 
Details of this method are discussed at length in D2K.
A slight modification from D2K is the following.
Since the galaxy does not
necessarily reside in the center of mass of the local
reference system, we apply a linear correction to account
for this displacement, where the gradients in the proper
motion in each coordinate are calculated from the 
proper motions of the local reference stars.
We found that this method gives a more robust
result than using a smaller number of local reference stars
(e.g., seven) and applying directly 
the median given by these stars to the motion of the galaxy.

In Figure 2 we show the proper motions in each coordinate
for the  galaxies 
as a function of $x$, $y$ and radius from the plate center.
Proper motion units are mas yr$^{-1}$ and throughout the
panels we use the same scale for comparison purposes. 
The filled symbols of the first and third row of panels
represent the proper motions as derived
from the central plate-overlap method (Section 2.2), i.e.,
the global solution. The open symbols (the second and fourth row of panels)
represent the proper
motions as derived from the local solution.
The error bars are the internal individual proper-motion errors
as determined in Section 2.2. The larger error bars
for galaxies situated at radii larger than 60 mm are
due to the use of a shorter time baseline; these images were 
not measurable on the Hale plates.
Clear trends with positions can be seen in the global solution
and, consequently, we obtain the large scatter in the outer regions.
Within a 40-mm radius we find
an $x$ proper-motion gradient of $\sim 0.05$ mas yr$^{-1}$ mm$^{-1}$
and a $y$ proper-motion gradient of $\sim 0.09$ mas yr$^{-1}$ mm$^{-1}$
for the global solution.
These gradients, over a distance of 20 mm, can easily produce a systematic
shift of 1 to 2 mas yr$^{-1}$. For reference, the majority of the 
cluster stars are within a 20-mm radius.
The local solution shows a significant improvement in terms of 
positional trends and scatter. 

Assured that our galaxy proper motions are free of systematics
we determine the mean correction to
absolute proper motion --- also called the zero point --- 
as a weighted mean of 
the proper motions derived from the local solution. The weights
are given by the individual proper-motion errors, (Section 2.2).
The uncertainty in the zero point was
calculated based on the scatter about the average and the  weights.

Fig. 2 shows that galaxies that otherwise
would cause a larger uncertainty in the solution
are given an appropriately lower weight. 
In Figure 3 we also show the proper motions of galaxies 
as a function of magnitude and color. No significant trends are 
visible. The QSO is highlighted with  a star symbol.

Our zero point with respect to field stars
within $ 16 \le V \le 19$ is  $\mu_{G,x}^F = 1.99 \pm 0.31$ mas yr$^{-1}$ and
$\mu_{G,y}^F = 4.83 \pm 0.37$ mas yr$^{-1}$.

\subsection{The Relative Mean Motion of the Cluster}

The mean motion of the cluster is determined by fitting in each coordinate
the sum of two Gaussians (that represent the cluster and the field)
to the proper-motion distribution
(e.g., Dinescu {\it et al.} 1996). Since proper-motion
errors increase quickly with magnitude at the faint end
(Fig. 1), we have restricted our sample to 
$ 16 \le V \le 20$.  Also, due to position-dependent systematics
discussed in the previous Section, we restrict the surveyed area 
for the relative motion of the cluster
to a $3\farcm64$ (20 mm) radius circle, centered on the plate center, which
roughly coincides with the cluster center.
Proper-motions larger than 15 mas yr$^{-1}$ in absolute value
are also discarded from the proper-motion distribution to be
modeled. In Figure 4 we present the observed, marginal
proper-motion distributions
(dotted line), smoothed by the individual proper-motion error
(see details in Girard {\it et al.} 1989) 
for a total of 405 stars, together with 
the two-component Gaussian least-squares, best-fit model (solid line).
In Table 2 we summarize all of the parameters from the best fit in each
coordinate. The numbers in parentheses are formal estimates of the
uncertainties as obtained from the fitting technique.
The fitted parameters along each axis are: 
the number of cluster stars, the center  
and the dispersion of the cluster proper-motion distribution,
and the center and the dispersion of the field
proper-motion distribution. From these parameters, the ratio of
cluster to field stars in the surveyed area is 1.55; in other words the
cluster population dominates this area.

The mean relative cluster motion is taken to be the center
of the cluster proper-motion distribution, and it is:
$\mu_{C,x}^R = 0.28 \pm 0.03$ mas yr$^{-1}$ and
$\mu_{C,y}^R = 0.34 \pm 0.05$ mas yr$^{-1}$.
The uncertainty in the mean cluster proper motion
is obtained by dividing the dispersion of the cluster distribution by the 
square root of the total number of cluster stars which is taken to be  
the average number from the $x$ and $y$ fit. This is a more realistic estimate
of the uncertainty in the mean motion compared to that given by
the formal estimate obtained from
the fit (Table 2).

Cluster membership probabilities based on the proper motions were also 
determined using the method from Dinescu {\it et al.} (1996)\footnote[8]
{A catalog of relative positions and proper motions, membership probabilities,
and photographic photometry is available via e-mail, from the first
author.}.

\subsection{Final Absolute Proper Motion}

The value of the cluster motion derived in the previous Section is
with respect to a mean reference system comprised of mainly
cluster stars but also field stars (the central-plate overlap
or global solution described in Section 2.2).
The reflex motion of the galaxies from the local solution
is derived with respect to field stars in the magnitude range
$ 16 \le V \le 19$.
We must therefore accurately determine the (median) motion of this
field-star component in our global-solution  proper-motion system 
which is admittedly affected by position-dependent systematics.

To derive the field motion, with an adequate accuracy, requires the use
of a larger number of field stars than what is available in the
immediate vicinity of the cluster, i.e., the area over which the
position-dependent systematics can be safely ignored.  Thus, we include
field stars up to 50 mm from the plate center and fit their proper motions
as a function of radius using an even-term polynomial of fourth order,
modeling out the systematic effects.
The magnitude range of the sample is widened slightly, to
$ 16 \le V \le 20$, to provide more stars.  (The gradient with magnitude
of the mean proper motion of the field is small enough to safely allow this.)
More importantly, the sample has been cleaned of contamination by the
cluster by eliminating all stars with proper motions within 1.4 mas yr$^{-1}$
of the mean motion of the cluster.
The radius of this
circle was chosen to be $2~\times$ max($\sigma_{C,x}, \sigma_{C,y}$).
A further cut in $x,y$ space, namely stars within 5 mm of the cluster, 
was also made to remove an overdensity that remained after the proper-motion
cut.  Finally, proper-motion outliers were discarded; those stars whose
motions in either coordinate exceeded 20 mas yr$^{-1}$.

The field star sample, selected in this manner, consisted of approximately
1100 stars.  The polynomial fits yielded a proper motion for the field,
at the position of the cluster, of
$\mu_{F,x}^{R} = -0.75 \pm 0.25$ mas yr$^{-1}$, and 
$\mu_{F,y}^{R} = -3.35 \pm 0.24$ mas yr$^{-1}$.
The fourth-order fit provided a smooth representation of the median, as
verified by overplotting the two.  The standard error of the fits, in
both the $x$ and $y$ components of $\mu_{F}^{R}$, was 4.6 mas yr$^{-1}$.

A subtle correction must be applied to the uncertainties associated with 
this derived motion for the field.  Approximately 400 of the field stars
are the same as those used in the local solution for the galaxy motion,
i.e., zero-point.  Thus, only 700 of the 1100 stars in this field sample
are drawn independently from the general field population of stars.
The error associated with any offset between
the average motion of the 400 stars in common and that of the general
field population will exactly cancel out when $\mu_{F,x,y}^{R}$ is combined
with $\mu_{G,x,y}^{F}$.  The appropriate additional uncertainty associated
with $\mu_{F,x,y}^{R}$, when it is being combined with $\mu_{G,x,y}^{F}$,
is 700/1100 of the formal values quoted above.

With this in mind, the absolute proper motion of NGC 7006 is
$\mu_{C,x,y} = \mu_{C,x,y}^{R} - \mu_{F,x,y}^{R} - \mu_{G,x,y}^{F}$.
We obtain $\mu_{C,x} = -0.96 \pm 0.35$ mas yr$^{-1}$,
and $\mu_{C,y} = -1.14 \pm 0.40$ mas yr$^{-1}$.

For comparison, had we used only two galaxies which are within 
a $4\arcmin$-radius (22 mm) from the plate center, 
(the area that we believe is only negligibly affected by 
position-dependent systematics)
and the global solution, we would have obtained the less accurate
values of $\mu_{C,x} = -1.28 \pm 0.48$ mas yr$^{-1}$,
and $\mu_{C,y} = -1.43 \pm 0.62$ mas yr$^{-1}$.  We adopt the former values,
based on the local solution above, as our best determination of the cluster's
absolute motion.

\section{The Orbit of NGC 7006}

The standard solar motion with respect to the Local
Standard of Rest (LSR) adopted here is 
$(U_{\odot}, V_{\odot}, W_{\odot}) = (-11.0, 14.0, 7.5)$ km s$^{-1}$
(Ratnatunga, Bahcall \& Casertano 1989). The $U$ component is positive 
outward from the GC. The adopted rotation velocity of
 the LSR is $\Theta_{0} = 220.0 $ km s$^{-1}$, and the solar circle radius is
8.0 kpc. The heliocentric distance to NGC 7006 is $40 \pm 2 $ kpc
(B91), and  the heliocentric
radial velocity is $-384.1 \pm 0.4$ km s$^{-1}$; 
the Galactic coordinates are $l = 63\fdg77$, $b = -19\fdg41$ (H96).
With the absolute proper motion  derived in Section 3.3 we obtain
the LSR velocity $(U, V, W) = (-116\pm62, -436\pm35, 149\pm60)$
 km s$^{-1}$.
The corresponding velocity in a cylindrical coordinate
system centered on the GC is $(\Pi,\Theta, W) = 
(-179\pm41, 155\pm66, 147\pm66) $ km s$^{-1}$.
In this left-handed system $\Pi$ is positive  outward from the Galactic 
rotation axis and $\Theta$ is positive in the direction of Galactic rotation,
both as seen at the cluster. The local circular velocity has been removed, leaving these velocities in the Galactic rest frame.

With the initial position and velocity, we have integrated the 
orbit of NGC 7006 in a three-component, analytical model of the
Galactic gravitational potential.
The bulge is represented by a Plummer potential, the disk by a 
Miyamoto \& Nagai (1975) potential, and the dark halo has a logarithmic
form. For the exact form of the potential see 
Paczy\'{n}ski (1990).

The orbital elements were calculated as in Dinescu {\it et al.} (1999b)
(hereafter DGvA99).
They are averages over a 10-Gyr time interval. The uncertainties 
in the orbital elements
were derived from the width of the distributions of orbital elements
over repeated integrations with different initial positions and
velocities. These positions and velocities were derived in a Monte Carlo 
fashion from the uncertainties in the observed quantities:
proper motions, distance and radial velocity. 
We obtain 
an orbit of pericentric radius $R_{p} = 17 \pm 4$ kpc, 
apocentric radius $R_{a} = 102 \pm 28$ kpc, maximum distance above the 
Galactic plane $z_{max} = 33 \pm 12$ kpc, eccentricity $e = 0.71 \pm0.02$, and
inclination with respect to the Galactic plane $\Psi = 26  \pm 9 \deg$.
The azimuthal period is $P_{\varphi} = (2.1 \pm 0.6) \times 10^{9}$ yr.
With the present location of NGC 7006 at a distance from the
GC of $\sim 37$ kpc and a distance of 13 kpc below the Galactic plane,
the cluster is presently moving toward the Galactic plane 
and toward the Galactic center.

\section{Discussion}

\subsection{Orbit Types of Globular Clusters and Galactic Satellites}

We proceed now to compare the orbital parameters of NGC 7006 with those
of the other globular clusters and Galactic satellites with well-determined
absolute proper motions.
The data for globulars are from
DGvA99 for the majority of the clusters,
D2K  for Pal 12, and Siegel {\it et al.} (2001) for Pal 13.
From the DGvA99 sample of clusters we have not included in the
present analysis Pal 3, a very sparse and
distant cluster (82 kpc, H96) whose absolute proper-motion
determination is a very challenging measurement.
The preliminary result (Majewski \& Cudworth 1993) is likely 
to be revised as a more extensive, improved study is
underway (Cudworth, private communication). 
Also, we have used revised heliocentric distances for clusters 
from H96, rather than the distances used in DGvA99.
Globular clusters considered here are all within 40 kpc of the GC.

The Galactic satellites included in this study are: the Large
Magellanic Cloud (LMC) and the dwarf spheroidals
 Sagittarius (Sgr), Ursa Minor (UMi) and Sculptor (Scl).  
The proper motion for the LMC is an average of three studies:
Jones, Klemola, \& Lin (1994), Kroupa, R$\ddot{o}$ser \& Bastian (1994),
and Kroupa \& Bastian (1997). 
The adopted heliocentric distance to the
LMC is $49\pm5$ kpc, and the heliocentric radial 
velocity is $270\pm4$ km s$^{-1}$ (e.g., Kroupa \& Bastian 1997,
Meatheringham {\it et al.} 1988).
Two proper-motion determinations are presented for Sgr:
the one derived by Irwin {\it et al.} (1996) from Schmidt 
plates (Sgr1), and the one derived by Ibata {\it et al.} (1998a) from HST WFPC2
frames (Sgr2), and quoted in Irwin (1998).
For Scl we have used the proper motion from Schweitzer {\it et al.} (1995)
(($\mu_{\alpha}$ cos $\delta$, $\mu_{\delta}$) = (0.72, -0.06) $\pm$ (0.22, 0.25)
mas yr$^{-1}$),  
and for UMi we used the proper-motion
determination from Schweitzer {\it et al.} (2001)
(($\mu_{\alpha}$ cos $\delta$, $\mu_{\delta}$) = 
(0.056, 0.078) $\pm$ (0.078, 0.099) mas yr$^{-1}$).
Positions, heloicentric distances and heliocentric radial velocities for
the dSphs were taken from the compilation of Mateo (1998). 

Orbital parameters were determined as in DGvA99,
in the potential model from Section 4.
The orbital elements for clusters of interest and satellites
are summarized in Table 3.
Uncertainties in the orbital parameters were also determined in a
Monte-Carlo fashion (DGvA99); however, we chose not to display them in the 
following Figure
because of the large range in their values from the low-energy to the
high-energy domain. This is mainly because of the uncertainty in the
heliocentric distance, which was chosen to be 10$\%$ of the 
distance (see DGvA99). Hence,  clusters at large distances will
have large uncertainties in their velocities and thereof in the
total energy and orbital angular momentum.
Uncertainties in the orbital parameters should however be kept in mind
when a dynamical association is considered. We have listed the uncertainties
in the orbital energy and orbital angular momentum for the
objects of interest in Table 3. These values 
are indicative of how susceptible to uncertainties  the apparent
phase-space associations are.

In Figure 5, top panel,  we show the total angular momentum $L$
as a function of the total orbital energy, $E$, for clusters that have 
$E > -10^{5}$ km$^2$ s$^{-2}$. $L$ was calculated as an
average over the entire integration time (10 Gyr); while it is
 not a strictly conserved quantity for the potential we have used, it
does provide some physical insight for the orbits
as it can be thought of as the third integral of motion. This is particularly
applicable for high-energy orbits, where the potential becomes more spherical
(see discussion in Binney \& Tremaine 1987).

The Galactic satellites and clusters of interest are labeled.
Clusters are represented with open squares, and satellites with
filled triangles. 
The units for energy are $10^4$ km$^2$ s$^{-2}$ and for
angular momentum are $10^4$ kpc km s$^{-1}$.
Among the clusters with measured absolute proper motions that are located
within 40 kpc from the GC, NGC 7006 is the most
energetic.   The other three clusters that fall in the same
category ($R_{a} > 80$ kpc, Table 3) are NGC 5466, NGC 6934 and Pal 13.
None of these four most energetic clusters seem to match the large total
angular momentum of the satellites at the same value of 
total orbital energy. 

In the middle panel we show the orbital
eccentricity as a function of the total orbital energy, for the entire 
energy range as defined by all globular
clusters and the four Galactic satellites. 
In the low-energy domain ($E < -10^{5}$ km$^2$ s$^{-2}$),
the distribution of eccentricities is rather uniform, while at
larger values of the orbital energy the orbits of the clusters
are preferentially of higher eccentricity. At the upper limit
of the orbital-energy domain, the clusters form a distinct population
that has highly eccentric orbits as opposed to the Galactic satellites
that have orbits with moderate eccentricities. 
From the inspection of Table 3, the four most energetic globulars 
have highly-eccentric orbits, with a large range in the inclination 
with respect to the Galactic plane (also called plunging orbits), 
while the Galactic satellites have more circular
orbits that are highly-inclined (also called polar orbits).
From the data available so far, the orbits
of outer halo clusters seem to be fundamentally different from those
of the Galactic satellites.

In order to inspect whether a possible dynamical association of NGC 7006
with other clusters and any of the Galactic satellites considered 
is apparent, we also plot the total orbital energy
as a function of the orbital angular momentum, $L_{z}$, for 
$E > -10^{5}$ km$^2$ s$^{-2}$ in the
bottom panel of Fig. 5.
From this plot, and from the previous plots in Fig. 5, it is apparent that
only NGC 5466 and NGC 6934 have similar orbital elements, while Pal 13
and NGC 7006 stand alone in the phase space. Also, none of the
four clusters considered here show any association with the Galactic
satellites considered, if similar values of the integrals of
motion are taken as evidence for a dynamical association. 
The association of NGC 5466 with NGC 6934 seen here was not
remarked upon in DGvA99, although the same values of the proper motions
were used. This new finding is due to the new heliocentric 
distances from H96 used here. However, the changes in the
orbital elements caused by the new heliocentric distances are smaller 
than the uncertainties in these elements as derived in DGvA99.
Therefore, this association should be regarded  as tentative because
of the large uncertainties in the orbits (see also Table 3).

We have marked two other clusters in Fig. 5, Pal 12 and
NGC 5024 (M 53). A special note is to be made regarding these clusters.
They share about the same locus as Sgr in all three plots. 
This is a clear indication of a common origin (see also 
Palma {\it et al.} 2001).
Pal 12's case has been extensively analyzed in D2K, where
strong evidence for tidal capture from Sgr was presented.
As for NGC 5024 --- from the plots presented above ---
exploring such a scenario may be a well-justified exercise.
However, we do not consider such an exercise in this work for two 
reasons: first, it would require a lengthy analysis that is
beyond the scope of this paper, and second, the reliability
of the analysis would be undermined by the proper-motion determination,
which has too large of an uncertainty for this purpose.
NGC 5024 has an absolute proper-motion determination 
which is subject to errors 
 of $\sim 1$ mas yr$^{-1}$ in each coordinate, while the size of the
proper-motion is a few tenths of mas yr$^{-1}$
(Odenkirchen {\it et al.} 1997). At an 18-kpc distance (H96) 
changes of the order of 1 mas yr$^{-1}$ 
can significantly alter the orbit.
In addition, the calibration to absolute proper motion
is not with respect to extragalactic objects, but to a few
$Hipparcos$ stars.  Usually these stars are much brighter 
(6 magnitudes) than cluster stars, and magnitude-dependent systematics 
are a major concern in photographic proper-motion studies
(e.g., Girard {\it et al.} 1998, Platais {\it et al.} 1998,
Dinescu {\it et al.} 1999a).
We strongly encourage a possibly new and/or improved
absolute proper-motion determination for NGC 5024.

Another note is required in this discussion. The Draco dSph
also has an absolute proper-motion determination (Scholz \& Irwin 1994).
The proper-motion uncertainty in each coordinate is of the order of
half a mas yr$^{-1}$. For a distant object
(80 kpc), such a determination  makes the orbit interpretation 
very uncertain. If considered however, Draco would have a highly eccentric
orbit in our infinite-mass potential;
in a more realistic potential it escapes the Galaxy.
Similarly, the LMC has a recent absolute proper-motion determination
based on QSOs and a short  time-baseline (8 years, Anguita {\it et al.} 2000).
This determination  disagrees with the more traditional 
determinations in the sense that the orbit is a lot more energetic. 
If this motion is correct, the LMC would have a highly eccentric orbit
or escape from the Galaxy, depending on the model adopted for the 
potential.
Since we feel that these results need confirmation 
from other studies, we chose not to interpret them at this point.

Although there is still little complete kinematical data on
outer halo globulars (R$_{GC} > 15$ kpc) and Galactic satellites, 
a picture emerges from what we know so far. If indeed outer clusters
have formed in individual, isolated, relatively low-mass systems
(often referred to as proto-Galactic fragments or building
blocks) that later were assimilated by the Milky Way
(Searle \& Zinn 1978, Zinn 1993, van den Bergh 2000 and
references therein), 
then it is apparent that the orbits of these systems
are quite different from those of the present-day Galactic satellites.
It may be that the character of the orbit played a major role in the
likelihood of survival of the satellite. 
Orbits that take satellites well into 
the inner regions of the Galaxy have a higher chance to be destroyed
because the local Galactic density becomes comparable to the central 
density of the satellite, a condition that initiates tidal disruption.
Models of satellite disruption show that the most dramatic 
effects take place during pericentric passages, when satellites can
lose as much as a third of their mass
(e.g., Johnston {\it et al.} 1999, Helmi 2000).
For realistic satellites and host galaxies 
(Helmi 2000, also Bassino {\it et al.} 1994) models show that orbits with
pericenters smaller than 20 kpc are highly destructive,
with dissolution timescales of 2-3 radial periods; 
for our cases, about 4-6 Gyr.

The satellites that are present today therefore can be thought of
as the survivors of a  system undergoing preferential destruction.
The main factors that contributed to the destruction are
orbit shape and satellite mass. For massive systems
( M $\ge  10^{10} $ M$_{\odot}$), dynamical friction plays a major role 
as the satellite loses orbital energy and spirals into the denser, inner
region of the Galaxy and
subsequently suffers destruction (e.g., Walker, Mihos \& Hernquist 1996).
One such case could very well be the Magellanic
Clouds (MC). According to the Murai \& Fujimoto (1980) model,
the apocentric distance for the MC has decreased by $50\%$ in the
past $10^{10}$ yr, placing the Clouds at a starting maximum distance of
$\sim 200$ kpc. However, present-day satellites are low-mass 
systems ($\sim 10^{7}$ M$_{\odot}$, Irwin \& Hatzidimitriou 1995). 
Only Fornax dSph
and Sgr are somewhat more massive; they are also the only dSphs known to
have their  own globular-cluster systems.
If the hypothetical parent satellites of the
clusters NGC 5466, NGC 6934 NGC 7006 and Pal 13
 had similar masses to Sgr and Fornax (up to
$10^{9}$ M$_{\odot}$),
then dynamical friction played a negligible role (e.g., Ibata \& Lewis 1998b),
and therefore the destruction is entirely due to the initial orbit
shape.

\subsection{Clues from Chemical Abundances}

The mass of a satellite plays an important role in its 
nucleosynthetic history: more massive satellites are able to retain
enriched gas from older generations of stars. Abundance patterns, and
in particular [$\alpha$/Fe] ratios, are powerful indicators
of the particulars of the star formation environment.
$\alpha$ elements are thought to be produced in type 
II supernovae, while type Ia supernovae produce mostly
iron-peak elements (Wheeler {\it et al.} 1989).
The traditional relationship of [$\alpha$/Fe] upon [Fe/H]
has a constant value of about 0.4 for $-2.0 \le $ [Fe/H] $ \le -1.0$,
and a gradual decline to 0.0 at [Fe/H] = 0.0. This is often 
interpreted as type II supernovae-dominated enrichment for 
[Fe/H] $ < -1$, followed by a gradually increasing contribution from 
type Ia supernovae as [Fe/H] increases. For example,
Shetrone, C\^{o}t\'{e} \& Sargent (2001) find that 
Draco, Sextans and Ursa Minor dwarfs 
have lower $\alpha$-enhancements than do the globular clusters
NGC 5272 (M 3), NGC 6341 (M 92) and NGC 2419, and halo field stars
for the same range in metallicity, $-3.0 < $ [Fe/H] $ < -1.2$
(see their Figure 4).
Specifically, Shetrone, C\^{o}t\'{e} \& Sargent (2001) determine 
that the three dSph  galaxies have 
$0.02 \le $ [$ \alpha$/Fe] $\le 0.13$ dex, while the three globulars have
a mean  of [$\alpha$/Fe] = $0.29 \pm 0.06$ dex, and halo field stars
have a mean of [$\alpha$/Fe] = $0.28 \pm 0.02$.
For NGC 7006, Kraft {\it et al.} (1998) find  [$ \alpha$/Fe] $\sim 0.3$
(see their Table 5).
McCarthy \& Nemec (1997) --- from the analysis of 
the anomalous Cepheid V19 in NGC 5466 --- find that the $\alpha$-ratio 
has a typical value for globular clusters, $\sim 0.3$ (see their Figure 11),
while NGC 6934 and Pal 13 have no such determinations to our knowledge.
Thus two of the four clusters of interest here match the 
abundance pattern of the majority of halo globulars (see also Carney 1996).
The lower $[\alpha$/Fe$]$ ratios for the relatively metal poor
dSphs implies that these systems either
lacked massive (M $> 10$ M$_{\odot}$) stars, or
were not able to retain ejecta from type II supernovae and thereby
incorporate these ejecta in the following generations of stars.
Therefore abundance patterns suggest that NGC 5466 and NGC 7006
could not have formed in environments of 
the type inferred for the low-mass dSphs Draco, Sextans and Ursa Minor.
In Sgr, however, most metal poor stars ([Fe/H] $\sim -1.5$)
 have [$\alpha$/Fe] similar
to typical halo stars and globulars
(Smecker-Hane \& McWilliam 1999), while [$\alpha$/Fe] for Fornax stars
has not been determined to date.
Therefore, the suggestion is that the hypothetical host Galactic satellites
of clusters NGC 5466 and NGC 7006 are more massive than the lower mass
(as opposed to the more massive dSphs Fornax and Sgr)
present-day dSphs.

We have also seen (Section 5.1) that the orbits of NGC 5466 and NGC 7006 are 
fundamentally different (plunging as opposed to polar orbits)
from those of the lower mass dSphs Ursa Minor and Sculptor.
Both chemical and dynamical arguments
suggest that at least some of the lower-mass dSphs may have a different 
formation history than the hypothetical parent fragments in which the 
outer halo globulars have formed.  One long-standing hypothesis
is that some dSphs formed as tidal condensations during the dynamical
interaction between our Galaxy and a massive satellite such as the LMC,
for instance. This scenario,
initially inspired by the spatial alignment of some dSphs along great circles
that include a more massive satellite (Kunkel \& Demers 1976,
 Lynden-Bell 1982, Majewski 1994), has recently gained more ground 
from a dynamical point of view. For instance, Ursa Minor's motion
(Schweitzer {\it et al.} 1997, their Figure 1, 
our Fig. 5 and Table 4) shows that the dSph is moving along the great circle
that contains the MC system and in the same sense as the MC system.
Olszewski (1998) points out however, that UMi is more metal poor than
the LMC and has a predominantly blue HB as opposed to the red HB of 
the LMC. Thus abundance arguments do not necessarily favor 
the tidal condensation scenario of UMi.

We turn now to investigate whether the hypothetical 
parent Galactic satellites of NGC 5466, NGC 6934, NGC 7006 
and Pal 13 could have 
resembled the more massive dSphs Sgr and Fornax by comparing 
the dSphs' cluster systems with the clusters under discussion.

\subsection{Comparison with Sgr and Fornax dSph Cluster Systems}

In Table 4 we summarize the metallicity, horizontal
branch (HB) type (B-V/B+V+R), absolute integrated magnitude M$_{V}$ 
and concentration parameters for NGC 5466, NGC 6934, NGC 7006 and Pal 13, 
for the clusters associated with Sgr (Da Costa \& Armandroff 1995, D2K),
and for the Fornax clusters. The data for the Galactic and
Sgr clusters are from H96, except for the concentration parameter
for Pal 12 which is from Rosenberg {\it et al.} (1998).
For the Fornax clusters, the metallicity and HB type are from
Buonanno {\it et al.} (1998, 1999), while the concentration
parameter and absolute integrated magnitude are from
Webbink (1985).

All four clusters considered here are known to be second-parameter
clusters (NGC 5466: Buonanno {\it et al.} 1985;
NGC 6934: Brocato {\it et al.} 1996,
NGC 7006: Sandage \& Wildey 1967, B91; Pal 13: e.g.,
Borissova {\it et al.} 1997).
Recent age determinations argue that 
NGC 5466, NGC 6943 and NGC 7006 are not younger than the bulk of the
globulars (see Rosenberg {\it et al.} 1999 for 
NGC 5466, Piotto {\it et al.} 1999 for NGC 6934, and
B91 for NGC 7006), while Pal 13's case may be somewhat uncertain
(Borissova {\it et al.} 1997).
Moreover, from an abundance analysis of NGC 7006 giants, Kraft {\it et al.}
 (1998) find that these stars have modest amounts of interior mixing, 
as opposed to stars in traditional clusters such as NGC 6752 and M13. 
Kraft {\it et al.} (1998) propose that this moderate amount of mixing 
may be responsible for the second parameter effect in this cluster as
opposed to age.
Another argument in favor of a canonical old age for NGC 7006 
is the ratio of $\alpha$ elements, which is similar to that of 
traditional globular clusters  (Kraft {\it et al.} 1998, Section 5.2).
Three of the Sgr clusters, namely  M54, Ter 8 and Arp 2 are also known
to have ages coeval with the rest of the halo globulars 
(Sarajedini \& Layden 2000); however these clusters have traditional
blue HB types for their metallicities, i.e. they do not display the
second parameter effect.    The other two Sgr clusters with red
HB types, Ter 7 and Pal 12 are metal richer ([Fe/H] $ > -1$), 
and younger by a few Gyr compared to traditional halo clusters
(for Ter 7 see Sarajedini \& Layden 2000 and references therein, for
Pal 12 see, e.g., Rosenberg {\it et. al} 1998).
The five Fornax
clusters are  known to display the second parameter effect 
(e.g., Buonanno 1998 and references therein). Recent HST-based age
determinations show that  the ages of Fornax clusters
are also coeval with those of halo globular clusters  
(Buonanno {\it et al.} 1998). One exception is the extreme second-parameter
cluster $\#4$, which has been determined to be significantly younger 
based on the HST-data analysis of the main sequence turnoff
(Buonanno {\it et al.} 1999).

Considering now only the coeval-age clusters, a sample that also includes
only clusters with [Fe/H] $ < -1.5$, we can see that 
NGC 5466, NGC 6934 and NGC 7006
have redder HB types than the Sgr clusters, and 
are more massive and more concentrated than these latter ones (Table 4).
One exception is M54, which is significantly more massive. M54 
has often been suggested to be the nucleus of Sgr (e.g., Larson 1996)
and can therefore be thought of as non-typical
for the globular cluster population.
However, NGC 5466, NGC 6934 and NGC 7006 have  
HB types, masses and concentrations that closely resemble
those of the Fornax cluster system (Table 4) rather than those of the Sgr 
system.

For the sake of completeness, we have also looked at the properties
of what now are known to be old globular clusters in the LMC. 
We refer the reader to the recent HST-based analysis of the 
CMDs of clusters in the LMC done by Olsen {\it et al.} (1998), and 
the discussion in their paper concerning ages and the references therein. 
From their Figure 17b, that
shows the relationship between metallicity and HB type, we can state that
our clusters in discussion, NGC 5466, NGC 6934 and NGC 7006, have HB types 
that are similar to those of the old LMC globular clusters, for the
corresponding metallicities.

Thus, at least three of the most energetic halo clusters known to date
have properties that resemble those of the cluster systems 
in Fornax and LMC rather than those of the Sgr cluster system.
It is also apparent that
the second parameter effect for clusters NGC 5466, NGC 6934 and NGC 7006,
 for four clusters in Fornax and possibly for some of those in the LMC,
is owed primarily to something other than age, perhaps related to
the environment where the clusters formed.

\subsection{Concluding Remarks}

The following facts now appear  to be secure for three of the most
energetic clusters known to date: they are fairly massive,
they formed in a type II supernovae-dominated environment
(except for NGC 6934 that has  no $\alpha$-ratio determination  to date),
they have highly eccentric orbits that are unlike the known orbits 
of the present-day satellites, and they display the second parameter effect
while having ages that are coeval with the ages of traditional,
first-parameter clusters.

We suggest that it is unlikely that NGC 5466, NGC 6934 and NGC 7006
 formed in very massive satellites
of the LMC type based on the highly eccentric character 
of the orbit (e $ \ge 0.7$, Table 3).
The N-body simulations of Tormen {\it et al.} (1998) that model the
survival of substructure in dark halos show that dynamical friction 
leads to some amount of orbital circularization. 
Therefore, had dynamical friction played a 
significant role in the history of the hypothetical parent Galactic satellites
of NGC 5466 and NGC 7006,
we would have expected moderate orbital eccentricities. In other words,
the initial orbit of the satellite should have been
altered by dynamical friction before the satellite was disrupted.
According to
the HB-type versus metallicity relation however, clusters NGC 5466, NGC 6934
 and NGC 7006 fit within the properties of the old LMC globulars.

It is more plausible that clusters such as NGC 5466, NGC 6934
and NGC 7006 formed
in satellites of the size of Sgr or Fornax that were completely destroyed
by tides rather early, owing to penetrating orbits into the 
denser regions of the Galaxy (Tormen {\it et al.} 1998). It is not clear 
however why the cluster properties such as mass and HB type of NGC 5466,
NGC 6934 and NGC 7006 better match those of the
Fornax clusters, rather than those of Sgr clusters.  
Based on the arguments discussed, we identify the Fornax dSph 
as highly representative of the Searle-Zinn proto Galactic fragments.
Fornax's survival may very well be due to its non-radial orbit.

Lastly, we note that, except for NGC 5466 and NGC 6934 which show 
a plausible dynamical asociation, NGC 7006 and Pal 13, 
stand alone in the phase space (Fig. 5).
Specifically, the large scatter in orbital angular momentum
(Fig. 5, bottom panel) shown for the ``pair'' NGC 5466-NGC 6934, NGC 7006 and
Pal 13 hints to a more chaotic assemblage of the
Milky Way outer halo, rather than the assemblage from the 
disruption of only one or two massive satellites.

\section{Summary}

We have measured the absolute proper motion of NGC 7006 and 
determined its orbit in a realistic Galactic potential.
Among the clusters with measured space velocities 
within 40 kpc from the GC, NGC 7006 is the most
energetic. 

We have compared the orbital characteristics of the clusters
NGC 5466, NGC 6934, NGC 7006 and Pal 13 --- which are the most 
energetic clusters known to date 
($R_{a} > 80$ kpc, see Table 3) ---
with those of satellite galaxies with well-measured proper motions.
We find no dynamical association of NGC 7006 with other clusters or 
Galactic satellites with well-measured, full space velocities.
This is also true for Pal 13. Only 
NGC 5466 and NGC 6934 show a possible common origin, as 
inferred from the integrals of motion, but no association with
the Galactic satellites considered here.

The common feature of the orbits of these four clusters is 
the ``orbit type'': 
highly eccentric, with various inclinations with respect to the
Galactic plane. This is in contrast with the orbits of the 
present-day Galactic satellites
which are of high inclination and small to moderate eccentricity.
We discuss possible causes for this difference, 
under the
assumption that outer halo clusters were formed in 
independently-evolving, proto galactic systems that were later 
assimilated by the Milky Way.  Specifically, one
hypothesis we set forth is that proto galactic fragments on
highly eccentric orbits that penetrated the denser regions of the 
Galaxy underwent dissolution rather early and quickly,
 leaving for the present day only those
systems on moderate eccentricities. Another explanation for the
orbit-type discrepancy between present-day Galactic satellites and
outer halo clusters is that 
some of the low-mass dSphs may have formed as condensations 
from the tidal interaction 
between a larger system, such as the MC and the Galaxy (e.g., Ursa Minor). 
Such dSphs have orbits that preserve the orbit type of the 
interacting system, i.e. the MC system that has a polar orbit.
Both these processes could have worked to produce the present Galactic outer 
halo and the Galactic satellite system. More kinematical data for
outer halo clusters and Galactic satellites would certainly help understand 
the formation picture of the halo. It would be very instructive to
learn whether the large scatter in the phase space, together with 
the plunging character of orbits, persists for the more remote halo clusters.

We have compared the properties of NGC 5466, NGC 6934, NGC 7006 
and Pal 13 with those of the 
clusters associated with Sgr and Fornax dwarf.
Based on masses, concentrations, and HB types we conclude that 
at least NGC 5466, NGC 6934 and NGC 7006 are more likely to have been
produced in Fornax-like systems.

We thank Allan Sandage and the Observatories of the Carnegie Institution 
for loan of the Hale 5 m plates taken by himself and W.~Baade. We are
grateful to Bill Schoening for providing us the KPNO 4 m plates.
We also wish to thank Tad Pryor, a diligent referee, whose thoughtful
comments helped us improve this final version of the paper.

This research was supported by NSF grant AST-97-02521.

\newpage

\begin{figure} 
\caption{Proper-motion errors and proper motions
as a function of magnitude (top and middle panel), and
proper motions as a function of color (bottom panel).
Filled circles represent cluster stars selected based on the CMD from
 B91. Proper-motion units throughout all of the figures are mas yr$^{-1}$.}
\end{figure}

\begin{figure}
\caption{Proper motions of extragalactic objects as a function of 
$x$ coordinate, $y$ coordinate, and radius from the plate center.
The central plate-overlap or global solution is represented 
with filled circles, while the local solution is
represented with open circles (see text for the description of
the proper-motion solutions).}
\end{figure}

\begin{figure}
\caption{Proper motions of extragalactic objects as a function of magnitude and
color, from the local solution. The QSO is highlighted with a 
star symbol.}
\end{figure}

\begin{figure}
\caption{Proper-motion marginal distributions along $\mu_x$ and $\mu_y$.
The dotted curves indicate the observed distributions, while the solid
curves show the least-squares, best-fit model to the observed
distributions by the sum of
two Gaussians, one representing the cluster, the other the field stars.}
\end{figure}

\begin{figure}
\caption{Angular momentum as a function of orbital energy (top panel),
eccentricity as a function of orbital energy (middle panel), and
orbital energy as a function of the $z$ component of the 
angular momentum, $L_{z}$
(bottom panel). Globular clusters are represented with open squares, while
the Galactic satellites are represented with filled triangles.
The clusters discussed in the text and the satellites are labeled.
The units for energy are $10^4$ km$^2$ s$^{-2}$ and for
angular momentum are $10^4$ kpc km s$^{-1}$.}
\end{figure}

\begin{table}[tbh] 
\begin{center}             
\begin{tabular}{lrrcl}                                 
\multicolumn{5}{c}{Table 1. Photographic Plates} \\ \\
 \hline\hline   \\
\multicolumn{1}{l}{Plate \#}&\multicolumn{1}{c}{Date} &
\multicolumn{1}{c}{H.A.}  &\multicolumn{1}{c}{Exp. }&      
\multicolumn{1}{c}{Emulsion+Filter} \\ 
& \multicolumn{1}{c}{(dd.mm.yy)} & & \multicolumn{1}{c}{(minutes)} & \\ 
\hline \\  
\multicolumn{5}{c}{Las Campanas DuPont 2.5 m (10\farcs92 mm$^{-1}$)} \\  \hline \\             
  CD3037 & 19.06.93 &  0.68 & 30 & IIa-O GG385 \\ 
  CD3038 & 19.06.93 & 23.98 & 30 & IIa-O GG385 \\ 
  CD3039 & 19.06.93 & 23.43 & 30 & IIa-O GG385 \\ 
  CD3048 & 20.06.93 &  0.37 & 45 & IIa-D GG495 \\ 
  CD3049 & 20.06.93 & 23.40 & 45 & IIa-D GG495 \\
  CD3056 & 21.06.93 &  0.22 & 30 & IIa-O GG385 \\ 
  CD3057 & 21.06.93 & 23.68 & 30 & IIa-O GG385 \\                     
  CD3058 & 21.06.93 & 23.12 & 30 & IIa-O GG385 \\ 
  CD3062 & 16.08.93 & 23.85 & 45 & IIa-O GG385 \\
  CD3098 & 21.08.93 &  0.30 & 45 & IIa-O GG385 \\ \hline \\
\multicolumn{5}{c}{KPNO 4 m (18\farcs6 mm$^{-1}$)} \\  \hline \\                  
  2874 & 27.08.78  &  0.88 & 40  &           IIIa-J GG385 \\
  3121 & 22.08.79  & 23.58 & 30  &           IIa-O GG385 \\
  3122 & 22.08.79  & 23.03 & 30  &           IIa-O GG385 \\
  3123 & 22.08.79  & 22.45 & 30  &           IIa-D GG495 \\
  3124 & 23.08.79  &  1.17 & 30  &           IIa-D GG495 \\
  3125 & 23.08.79  &  0.67 & 30  &           IIa-D GG495 \\
  3138 & 24.08.79  & 23.65 & 30  &           IIa-O GG385 \\ \hline \\
\multicolumn{5}{c}{Hale 5 m (11\farcs12 mm$^{-1}$)} \\  \hline \\        
PH808s & 02.10.54  & 0.17 &  10 & 103a-O GG13\\
PH809s & 02.10.54  & 23.87 &  10 & 103a-O GG13 \\
PH824s & 03.10.54  & 23.27 &  20 & 103a-D GG11 \\
PH1258s & 11.08.56  & 22.57 &  10 & 103a-D GG11 \\
PH1261s & 11.08.56  & 21.78 &  10 & 103a-D GG11 \\
PH1292s & 13.08.56  & 23.62 &   7 & 103a-D GG11 \\
PH1295s & 13.08.56  & 22.92 &   7 & 103a-D GG11 \\
PH1296s & 13.08.56  & 22.73 &   7 & 103a-D GG11 \\
\hline                                           
\end{tabular}                                                    
\end{center}                                                     
\end{table}

\begin{table}[tbh] 
\begin{center}             
\begin{tabular}{llllll}
\multicolumn{6}{c}{Table 2. Model Parameters } \\ \\
\hline\hline   \\
& \multicolumn{1}{c}{N$_C$} & \multicolumn{1}{c}{$\mu_C$} &
\multicolumn{1}{c}{$\sigma_C$} & \multicolumn{1}{c}{$\mu_F$} &
\multicolumn{1}{c}{$\sigma_F$} \\
& & \multicolumn{4}{c}{(mas yr$^{-1}$)} \\
\hline \\ 
\multicolumn{1}{c}{X:} & \multicolumn{1}{c}{243(3)} & \multicolumn{1}{c}{0.280(3)} & \multicolumn{1}{c}{0.464(4)}
& \multicolumn{1}{c}{-0.757(10)} & \multicolumn{1}{c}{2.919(10)} \\
\multicolumn{1}{c}{Y:} & \multicolumn{1}{c}{249(3)} & \multicolumn{1}{c}{0.343(5)} & \multicolumn{1}{c}{0.682(6)}
& \multicolumn{1}{c}{-2.168(13)} & \multicolumn{1}{c}{3.817(11)} \\
\hline \\
\end{tabular}
\end{center}
\end{table}

\begin{table}[tbh]
\begin{center}
\begin{tabular}{rrrrrrrrrr}
\multicolumn{9}{c}{Table 3. Orbital Elements} \\ \\
\hline
\hline
\\
\multicolumn{1}{c}{Object} & 
\multicolumn{1}{c}{$E$}&
\multicolumn{1}{c}{$L_{z}$}&
\multicolumn{1}{c}{$L$} &
\multicolumn{1}{c}{$P_{\varphi}$}& 
\multicolumn{1}{c}{$R_{a}$}&
\multicolumn{1}{c}{$R_{p}$}&
\multicolumn{1}{c}{$z_{max}$} &
\multicolumn{1}{c}{e} & 
\multicolumn{1}{c}{$\Psi$} \\
& \multicolumn{1}{c}{($10^{4}$km$^{2}$s$^{-2}$)} &
\multicolumn{2}{c}{(kpc kms$^{-1}$)} &
\multicolumn{1}{c}{($10^{9}$ yr)}& 
\multicolumn{1}{c}{(kpc)} &
\multicolumn{1}{c}{(kpc)} &
\multicolumn{1}{c}{(kpc)} &&
\multicolumn{1}{c}{($\arcdeg$)} \\ \\
\hline \\ 
\multicolumn{1}{l}{NGC 5466} & \multicolumn{1}{r}{-2.4(1.6)} &
\multicolumn{1}{r}{-610(287)} & \multicolumn{1}{r}{3715} &
\multicolumn{1}{r}{2.0} & 
\multicolumn{1}{r}{96} &
\multicolumn{1}{r}{9} & \multicolumn{1}{r}{54} &
\multicolumn{1}{r}{0.84} & \multicolumn{1}{r}{37} \\
\multicolumn{1}{l}{NGC 6934} & \multicolumn{1}{r}{-2.6(1.9)} &
\multicolumn{1}{r}{-54(623)} & \multicolumn{1}{r}{3766} &
\multicolumn{1}{r}{1.9} & 
\multicolumn{1}{r}{88} &
\multicolumn{1}{r}{9} & \multicolumn{1}{r}{56} &
\multicolumn{1}{r}{0.81} & \multicolumn{1}{r}{73} \\
\multicolumn{1}{l}{NGC 7006} & \multicolumn{1}{r}{-2.0(0.9)} &
\multicolumn{1}{r}{5420(1732)} & \multicolumn{1}{r}{6431} &
\multicolumn{1}{r}{2.1} & 
\multicolumn{1}{r}{102} &
\multicolumn{1}{r}{17} & \multicolumn{1}{r}{33} &
\multicolumn{1}{r}{0.71} & \multicolumn{1}{r}{26} \\
\multicolumn{1}{l}{Pal 13} & \multicolumn{1}{r}{-2.3(0.8)} &
\multicolumn{1}{r}{-3016(729)} & \multicolumn{1}{r}{4721} &
\multicolumn{1}{r}{2.0} & 
\multicolumn{1}{r}{96} &
\multicolumn{1}{r}{12} & \multicolumn{1}{r}{35} &
\multicolumn{1}{r}{0.78} & \multicolumn{1}{r}{30} \\
\hline \\
\multicolumn{1}{l}{Scl} & \multicolumn{1}{r}{-0.7(1.4)} &
\multicolumn{1}{r}{2038(606)} & \multicolumn{1}{r}{15858} &
\multicolumn{1}{r}{3.0} & 
\multicolumn{1}{r}{124} &
\multicolumn{1}{r}{61} & \multicolumn{1}{r}{91} &
\multicolumn{1}{r}{0.34} & \multicolumn{1}{r}{69} \\
\multicolumn{1}{l}{LMC} & \multicolumn{1}{r}{-2.1(1.3)} &
\multicolumn{1}{r}{2568(2405)} & \multicolumn{1}{r}{10820} &
\multicolumn{1}{r}{2.0} & 
\multicolumn{1}{r}{85} &
\multicolumn{1}{r}{41} & \multicolumn{1}{r}{60} &
\multicolumn{1}{r}{0.35} & \multicolumn{1}{r}{67} \\
\multicolumn{1}{l}{UMi} & \multicolumn{1}{r}{-1.5(0.4)} &
\multicolumn{1}{r}{3740(930)} & \multicolumn{1}{r}{13024} &
\multicolumn{1}{r}{2.7} & 
\multicolumn{1}{r}{98} &
\multicolumn{1}{r}{51} & \multicolumn{1}{r}{71} &
\multicolumn{1}{r}{0.32} & \multicolumn{1}{r}{66} \\
\multicolumn{1}{l}{Sgr1} & \multicolumn{1}{r}{-4.5(0.8)} &
\multicolumn{1}{r}{-108(233)} & \multicolumn{1}{r}{4357} &
\multicolumn{1}{r}{1.1} & 
\multicolumn{1}{r}{50} &
\multicolumn{1}{r}{13} & \multicolumn{1}{r}{31} &
\multicolumn{1}{r}{0.58} & \multicolumn{1}{r}{58} \\
\multicolumn{1}{l}{Sgr2} & \multicolumn{1}{r}{-4.6(1.9)} &
\multicolumn{1}{r}{827(986)} & \multicolumn{1}{r}{4294} &
\multicolumn{1}{r}{1.0} & 
\multicolumn{1}{r}{48} &
\multicolumn{1}{r}{13} & \multicolumn{1}{r}{29} &
\multicolumn{1}{r}{0.56} & \multicolumn{1}{r}{57} \\
\hline 
\end{tabular}
\end{center}
\end{table}

\begin{table}[tbh] 
\begin{center}             
\begin{tabular}{lrrrr}                                 
\multicolumn{5}{c}{Table 4. Cluster Parameters} \\ \\
 \hline\hline   \\
\multicolumn{1}{l}{Cluster}&\multicolumn{1}{r}{[Fe/H]} & 
\multicolumn{1}{c}{HB-type} &
\multicolumn{1}{c}{M$_{V}$}  &\multicolumn{1}{c}{c } \\ \\ 
 \hline \\
NGC 5466 & -2.22 &  0.58 & -7.11 & 1.32 \\
NGC 6934 & -1.54 &  0.25 & -7.65 & 1.53 \\
NGC 7006 & -1.63 & -0.28 & -7.68 & 1.42 \\
Pal 13   & -1.65 & -0.20 & -3.51 & 0.66 \\
 \hline \\
\multicolumn{5}{c}{Sgr clusters} \\ \\
  M 54 & -1.59 & 0.87 & -10.01 & 1.84 \\ 
 Ter 7 & -0.58 & -1.00 & -5.05 & 1.08 \\
 Arp 2 & -1.76 & 0.86 & -5.29 & 0.90 \\
 Ter 8 & -2.00 & 1.00 & -5.05 & 0.60 \\
 Pal 12& -0.94 & -1.00 & -4.48 & 1.08 \\
\hline                                            \\
\multicolumn{5}{c}{Fornax clusters} \\ \\
  $\#$1 & -2.20 & -0.20 & -5.23 & 0.71 \\ 
  $\#$2 & -1.79 &  0.38 & -7.30 & 1.08 \\
  $\#$3 (NGC 1049) &  -1.96 & 0.50 & -8.19 & 1.83 \\
  $\#$4 & -1.90 & -1.00 & -7.23 & 1.82 \\
  $\#$5 & -2.20 & 0.44 & -7.38 & 1.26 \\
\hline
\end{tabular}                                                    
\end{center}                                                     
\end{table}

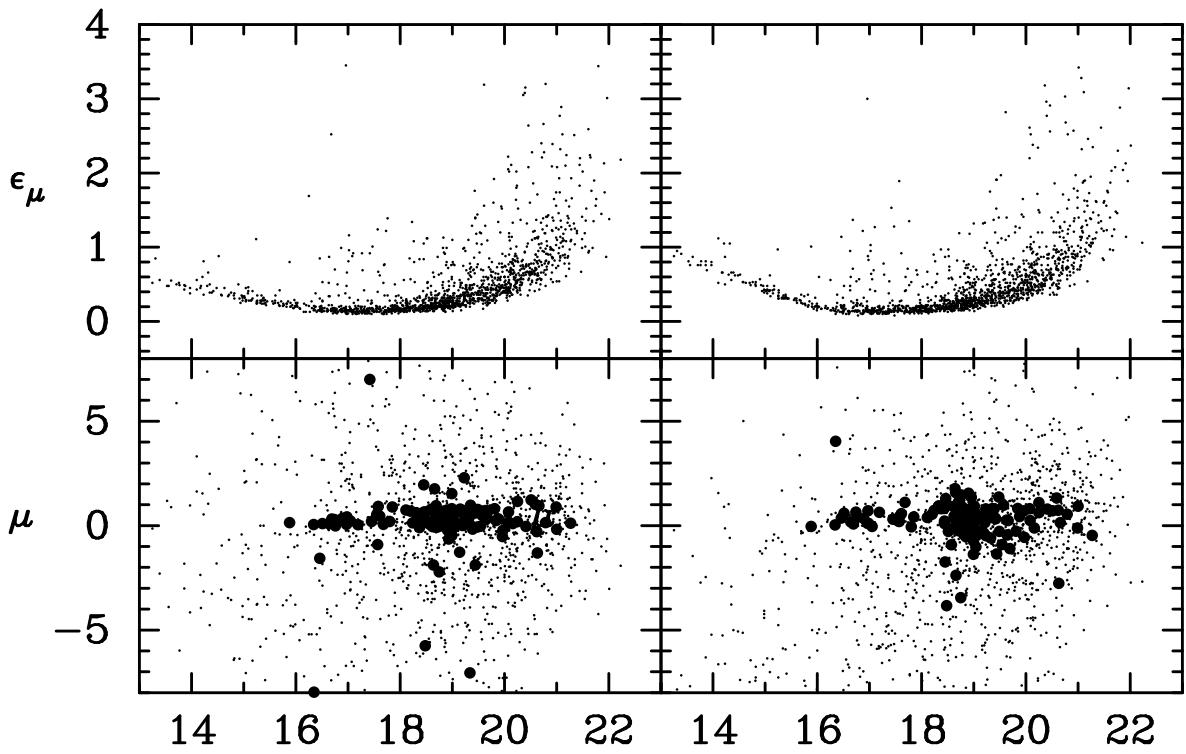
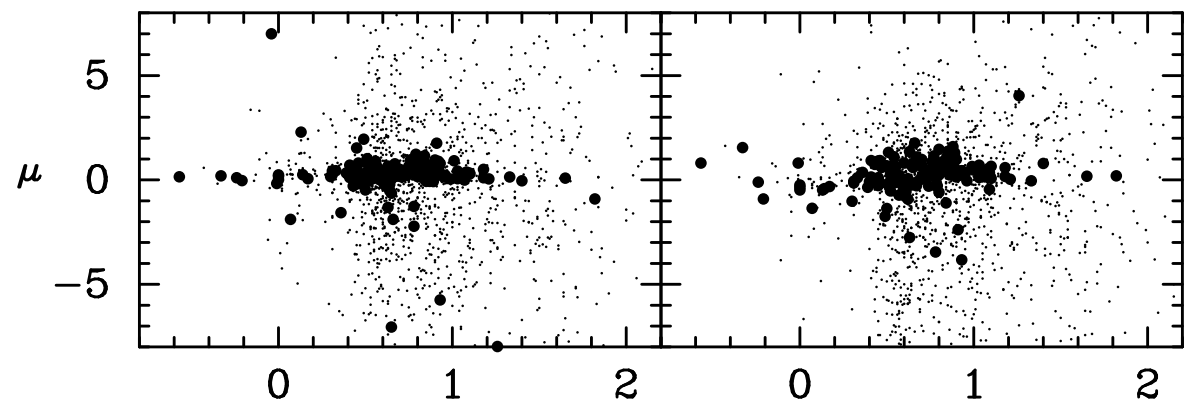
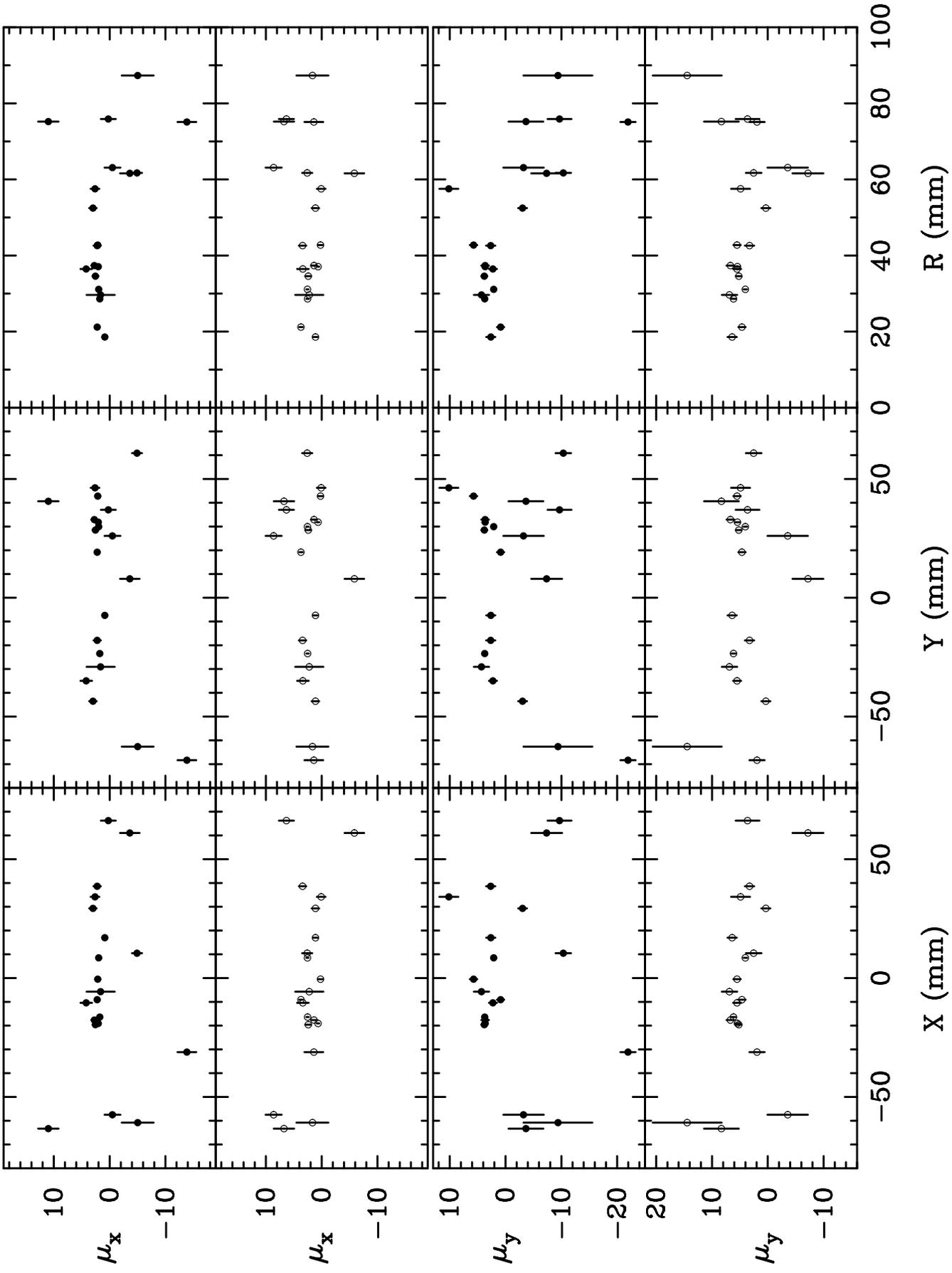
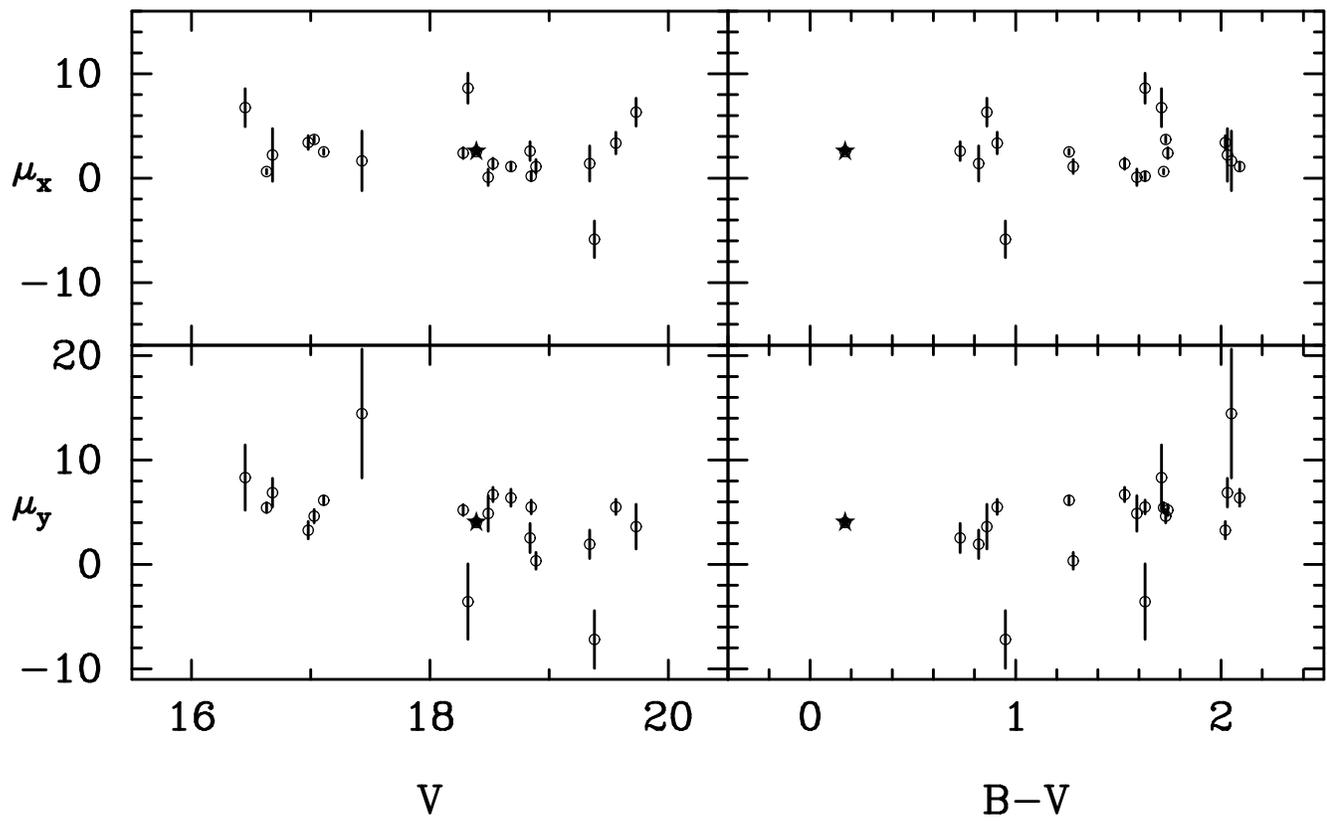
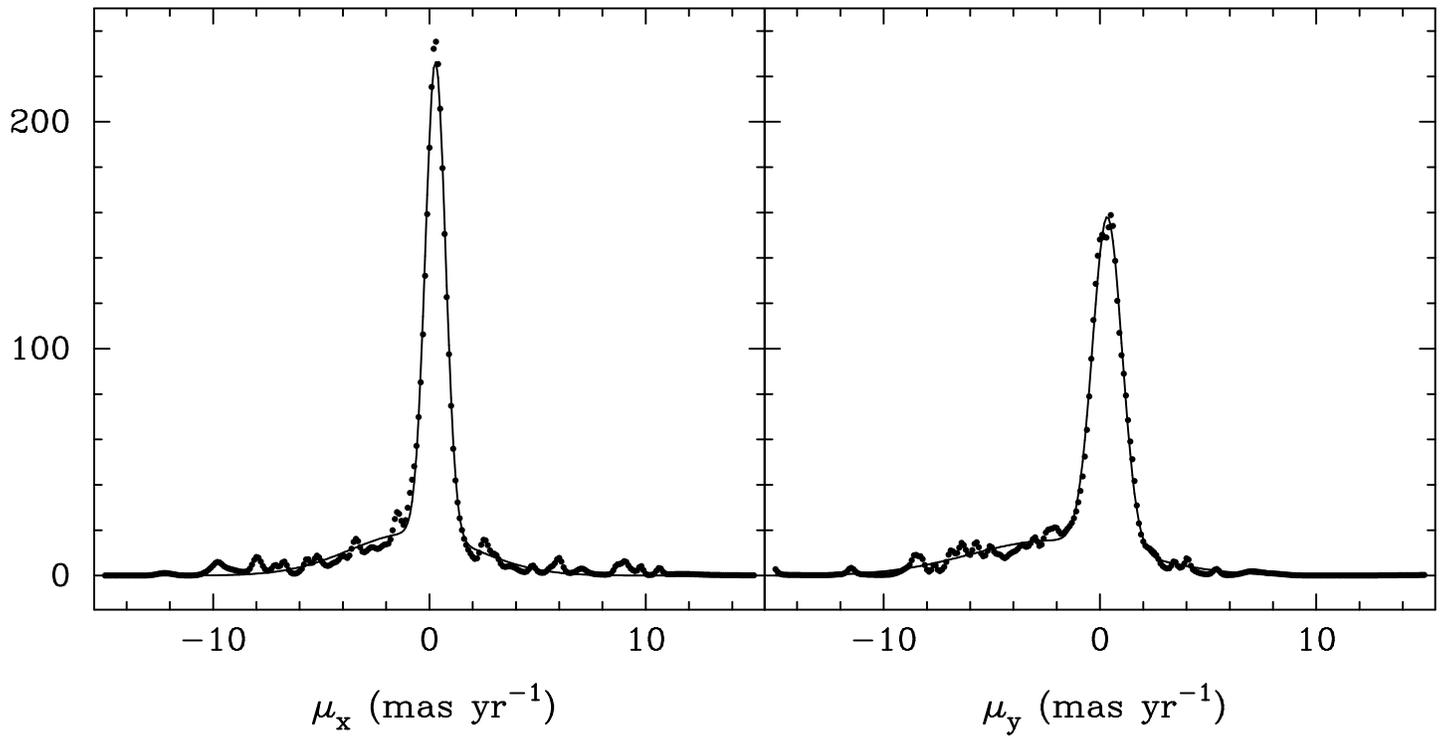
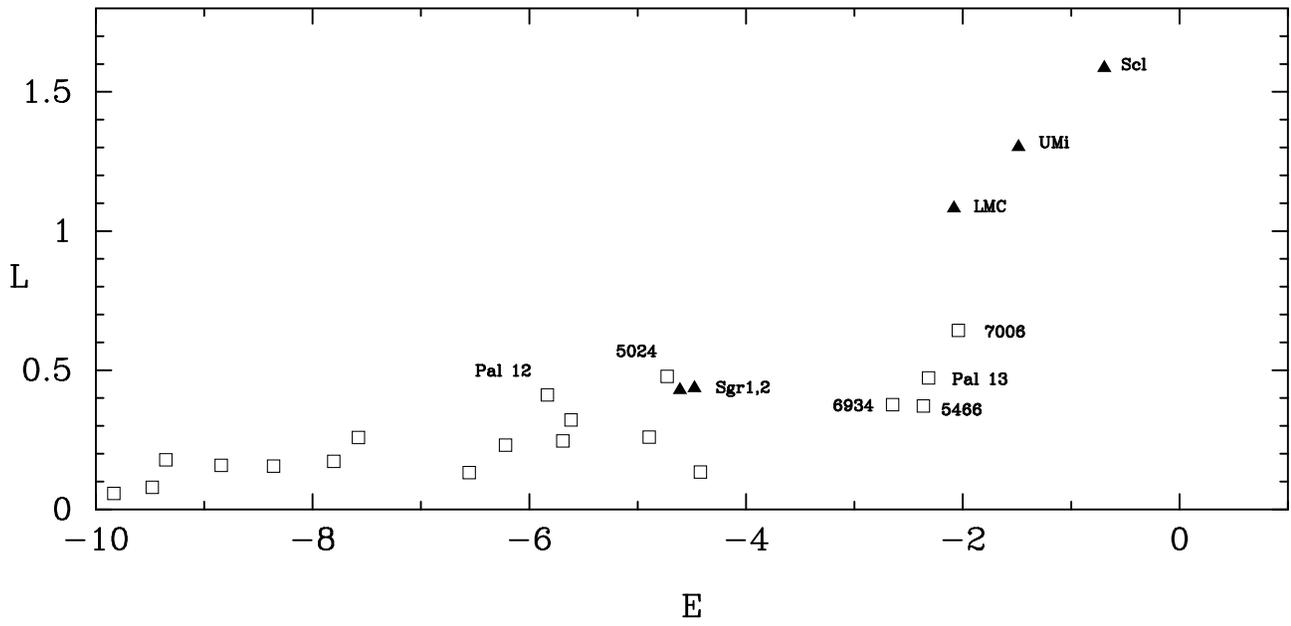
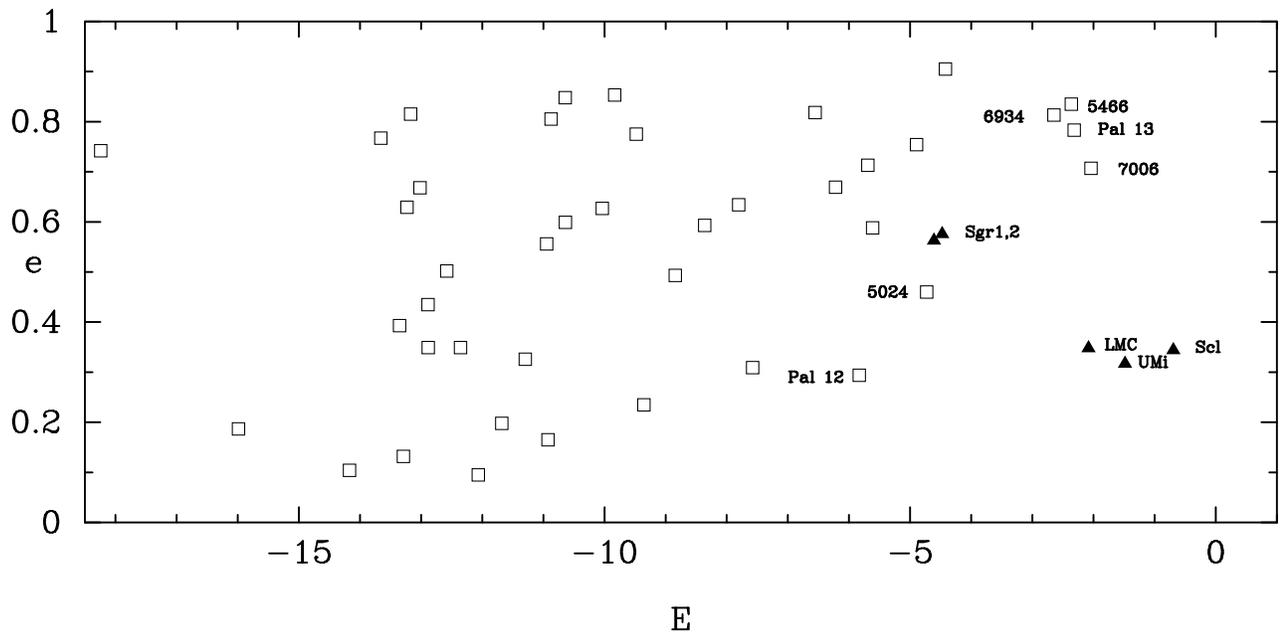
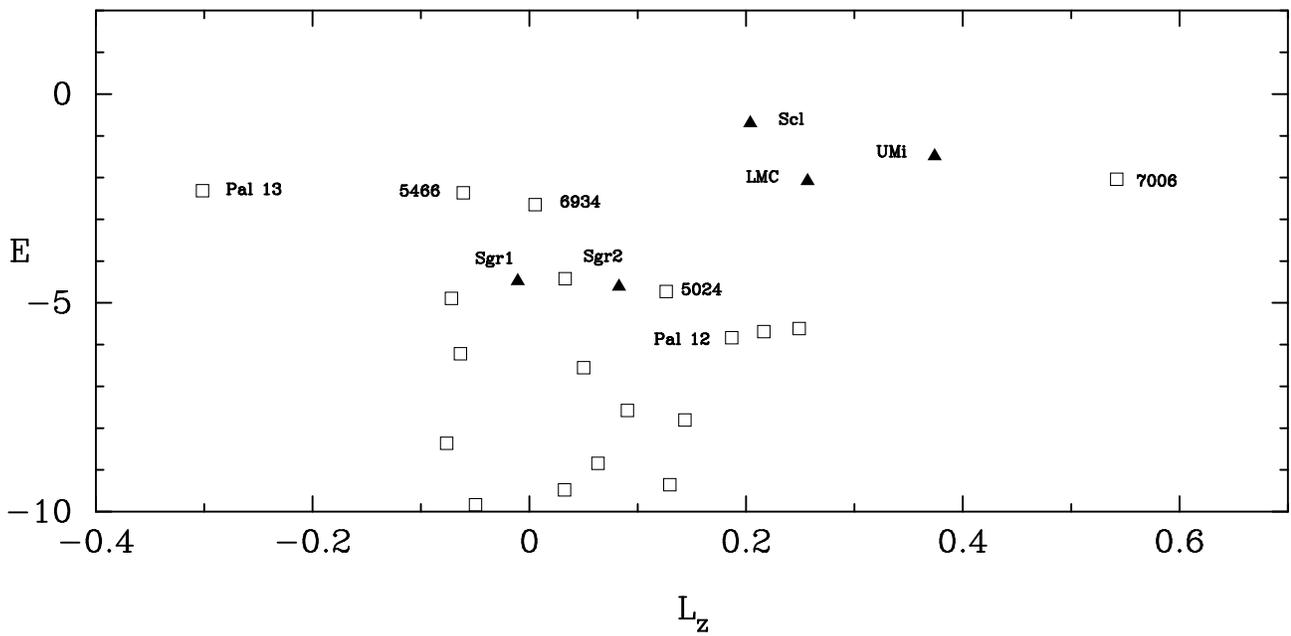
\end{document}